\documentclass[12pt]{article}
\usepackage{graphicx}
\usepackage{color}
\usepackage{authblk}
\def\hybrid{\topmargin 0pt      \oddsidemargin 0pt
        \headheight 0pt \headsep 0pt
       \voffset-1cm
        \textwidth 6.25in       % A4 paper
       \textheight 9.5in       % A4 paper
        \marginparwidth 0.0in
        \parskip 5pt plus 1pt   \jot = 1.5ex}
\catcode`\@=11
\def\marginnote#1{}

\newcount\hour
\newcount\minute
\newtoks\amorpm
\hour=\time\divide\hour by60
\minute=\time{\multiply\hour by60 \global\advance\minute by-\hour}
\edef\standardtime{{\ifnum\hour<12 \global\amorpm={am}%
        \else\global\amorpm={pm}\advance\hour by-12 \fi
        \ifnum\hour=0 \hour=12 \fi
        \number\hour:\ifnum\minute<10 0\fi\number\minute\the\amorpm}}
\edef\militarytime{\number\hour:\ifnum\minute<10 0\fi\number\minute}

\def\draftlabel#1{{\@bsphack\if@filesw {\let\thepage\relax
   \xdef\@gtempa{\write\@auxout{\string
      \newlabel{#1}{{\@currentlabel}{\thepage}}}}}\@gtempa
   \if@nobreak \ifvmode\nobreak\fi\fi\fi\@esphack}
        \gdef\@eqnlabel{#1}}
\def\@eqnlabel{}
\def\@vacuum{}
\def\draftmarginnote#1{\marginpar{\raggedright\scriptsize\tt#1}}

\def\draftlabel#1{{\@bsphack\if@filesw {\let\thepage\relax
   \xdef\@gtempa{\write\@auxout{\string
      \newlabel{#1}{{\@currentlabel}{\thepage}}}}}\@gtempa
   \if@nobreak \ifvmode\nobreak\fi\fi\fi\@esphack}
        \gdef\@eqnlabel{#1}}
\def\@eqnlabel{}
\def\@vacuum{}
\def\draftmarginnote#1{\marginpar{\raggedright\scriptsize\tt#1}}

\def\draft{\oddsidemargin -.5truein
        \def\@oddfoot{\sl preliminary draft \hfil
        \rm\thepage\hfil\sl\today\quad\militarytime}
        \let\@evenfoot\@oddfoot \overfullrule 3pt
        \let\label=\draftlabel
        \let\marginnote=\draftmarginnote
   \def\@eqnnum{(\theequation)\rlap{\kern\marginparsep\tt\@eqnlabel}%
\global\let\@eqnlabel\@vacuum}  }

\def\numberbysection{\@addtoreset{equation}{section}
        \def\theequation{\thesection.\arabic{equation}}}

\def\underline#1{\relax\ifmmode\@@underline#1\else
        $\@@underline{\hbox{#1}}$\relax\fi}

\def\titlepage{\@restonecolfalse\if@twocolumn\@restonecoltrue\onecolumn
     \else \newpage \fi \thispagestyle{empty}\c@page\z@
        \def\thefootnote{\fnsymbol{footnote}} }

\def\endtitlepage{\if@restonecol\twocolumn \else  \fi
        \def\thefootnote{\arabic{footnote}}
        \setcounter{footnote}{0}}  %\c@footnote\z@ }
\relax

\numberbysection
\hybrid

\newfont{\Bbb}{msbm10 scaled 1\@ptsize00}
\newfont{\Bbbb}{msbm7 scaled 1\@ptsize00}

\newcommand{\DDD}{\raise-1pt\hbox{$\mbox{\Bbbb D}$}}

\newcommand{\UUU}{\raise-1pt\hbox{$\mbox{\Bbbb U}$}}

\newcommand{\z}{\raise-1pt\hbox{$\mbox{\Bbbb Z}$}}

\def\res{\mathop{\hbox{res}}\limits}

\def\beq{\begin{equation}}
\def\eeq{\end{equation}}
\def\p{\partial}

\begin{document}

\begin{titlepage}

\title{Toda lattice hierarchy and trigonometric Ruijsenaars-Schneider hierarchy}

\author[1,2]{V.~Prokofev\thanks{vadim.prokofev@phystech.edu }}
\author[2,3,4]{
 A.~Zabrodin\thanks{ zabrodin@itep.ru}}
 \affil[1]{Moscow Institute of Physics and Technology, Dolgoprudny, Institutsky per., 9,
Moscow region, 141700, Russia}
 \affil[2]{
Skolkovo Institute of Science and Technology, 143026 Moscow, Russian Federation
}
\affil[3]{ National Research University Higher School of Economics,
20 Myasnitskaya Ulitsa, Moscow 101000, Russian Federation}
\affil[4]{
ITEP, 25 B.Cheremushkinskaya, Moscow 117218, Russian Federation
}

%\vspace{-5cm} \centerline{ \hfill ITEP-TH-17/19}\vspace{7cm}

\date{July 2019}
\maketitle

\vspace{-11cm} \centerline{ \hfill ITEP-TH-17/19}\vspace{11cm}

\begin{abstract}

We consider solutions of the 2D Toda lattice hierarchy which are trigonometric 
functions of the ``zeroth'' time $t_0=x$. It is known that their poles move as
particles of the trigonometric Ruijsenaars-Schneider model. We extend this correspondence
to the level of hierarchies: the dynamics of poles with respect to the $m$-th hierarchical
time $t_m$ (respectively, $\bar t_m$) of the 2D Toda lattice hierarchy is shown to be 
governed by the Hamiltonian which is proportional to
the $m$-th Hamiltonian $\mbox{tr}\, L^m$ (respectively, $\mbox{tr}\, L^{-m}$)
of the Ruijsenaars-Schneider model, where $L$ is the 
Lax matrix. 

\end{abstract}

\end{titlepage}

\tableofcontents

\vspace{5mm}

\section{Introduction}

The 2D Toda lattice (2DTL) hierarchy is an infinite set
of compatible nonlinear dif\-fe\-ren\-ti\-al-\-dif\-fe\-rence 
equations involving infinitely many
time variables ${\bf t}=\{t_1, t_2, t_3, \ldots \}$ (``positive'' times), 
$\bar {\bf t}=\{\bar t_1, \bar t_2, \bar t_3, \ldots \}$ (``negative'' times)
in which the equations
are differential and the ``zeroth'' time 
$t_0=x$ in which the equations are difference. When the negative times are frozen, 
the equations involving $x$ and ${\bf t}$ variables form the modified Kadomtsev-Petviashvili
(mKP) hierarchy which is a subhierarchy of the 2DTL one. 
Among all solutions to these equations,
of special interest are solutions which have a finite number of poles 
in the variable $x$ in a fundamental domain of 
the complex plane. In particular, one can consider solutions
which are trigonometric or hyperbolic functions of $x$ with poles depending on the times.

The investigation of dynamics of poles of singular solutions to nonlinear integrable
equations was initiated in the seminal paper \cite{AMM77}, where elliptic and rational 
solutions to the Korteweg-de Vries and Boussinesq equations were studied. It was shown
that the poles move as particles of the integrable Calogero-Moser many-body system
\cite{Calogero71,Calogero75,Moser75,OP81}
with some restrictions in the phase space. As it was proved in \cite{Krichever78,CC77},
this connection becomes most natural for the more general 
Kadomtsev-Petviashvili (KP) equation, in which case there are no restrictions in the phase space 
for the Calogero-Moser dynamics of poles. 
The method suggested by Krichever \cite{Krichever80} for elliptic solutions 
of the KP equation consists in substituting the solution not in the KP equation itself but
in the auxiliary linear problem for it (this implies a suitable pole ansatz for the wave
fuction). This method allows one to obtain the equations of motion together with their
Lax representation. 

The further progress was achieved in Shiota's work \cite{Shiota94}. Shiota has shown that 
the correspondence between rational solutions to the KP equation and the Calogero-Moser
system with rational potential can be extended to the level of hierarchies. 
The evolution of poles with respect to the higher times $t_k$ 
of the KP hierarchy was shown to be governed by 
the higher Hamiltonians $H_k=\mbox{tr}\, L^k$
of the integrable Calogero-Moser system, where $L$ is the Lax matrix. 
Later this correspondence was generalized to trigonometric
solutions of the KP hierarchy (see \cite{Haine07,Z19a}). 

Dynamics of poles of elliptic solutions to the 2DTL and 
mKP hierarchies was 
studied in \cite{KZ95}. It was proved that the poles move as particles of the 
integrable Ruijsenaars-Schneider many-body system \cite{RS86} 
which is a relativistic generalization
of the Calogero-Moser system. The extension to the level of hierarchies
for rational solutions to the mKP equation has been made 
in \cite{Iliev07} (see also \cite{Z14}): again, the 
evolution of poles with respect to the higher times $t_k$ 
of the mKP hierarchy is governed by 
the higher Hamiltonians $\mbox{tr}\, L^k$ of the Ruijsenaars-Schneider system. 

In this paper we study the correspondence of the 
2DTL hierarchy and the Ruijsenaars-Schneider
hierarchy for trigonometric solutions of the former. Our method consists in a direct
solution of the auxiliary linear problems for the wave function and its adjoint using a
suitable pole ansatz. The tau-function of the 2DTL hierarchy for trigonometric 
solutions has the form
$$
\tau (x, {\bf t}, \bar {\bf t})=\exp \Bigl (-\sum_{k\geq 1}kt_k \bar t_k \Bigr )
\prod_{i=1}^{N}\Bigl ( e^{2\gamma x}-
e^{2\gamma x_i({\bf t}, \bar {\bf t})}\Bigr ),
$$
where $\gamma$ is a complex parameter. (The zeros $x_i$ of the tau-function are poles of the 
solution.)
When $\gamma$ is purely imaginary (respectively, real),
one deals with trigonometric (respectively, hyperbolic) solutions. The limit $\gamma \to 0$
corresponds to rational solutions. We show that the evolution of the $x_i$'s with respect to the
time $t_m$ is governed by the Hamiltonian
\beq\label{i1}
H_m=-\, \frac{\sinh (m\gamma \eta )}{m\gamma \eta}\, \mbox{tr}\, L^m,
\eeq
where the parameter $\eta$ has the meaning of the inverse velocity of light and
\beq\label{i2}
L_{ij}=\frac{\gamma \eta \, e^{\eta p_i}}{\sinh (\gamma (x_i-x_j-\eta ))}
\prod_{l\neq i}\frac{\sinh (\gamma (x_i-x_l+\eta ))}{\sinh (\gamma (x_i-x_l))}
\eeq
is the Lax matrix of the trigonometric Ruijsenaars-Schneider system. 
In particular, 
\beq\label{i3}
H_1= \sum_i e^{\eta p_i}
\prod_{l\neq i}\frac{\sinh (\gamma (x_i-x_l+\eta ))}{\sinh (\gamma (x_i-x_l))}
\eeq
is the standard first Hamiltonian of the Ruijsenaars-Schneider system. In a similar way, 
the evolution of the $x_i$'s with respect to the time $\bar t_m$ is governed by the 
Hamiltonian
\beq\label{i1a}
\bar H_m=-\, \frac{\sinh (m\gamma \eta )}{m\gamma \eta}\, \mbox{tr}\, L^{-m}.
\eeq

The paper is organized as follows. In section 2 we remind the reader the main facts
about the 2DTL hierarchy. Section 3 is devoted to solutions which are trigonometric functions
of $x=t_0$. We derive equations of motion for their poles as functions of the time $t_1$.
In section 4 we consider the dynamics of poles with respect to the higher times and derive
the corresponding Hamiltonian equations. In section 5 we derive the 
self-dual form of equations of motion and show that it encodes 
all higher equations of motion in the generating form. 
In section 6 the determinant formula for the 
tau-function of trigonometric solutions is proved. 

\section{The 2D Toda latttice hierarchy}

Here we very briefly review the 2DTL hierarchy (see \cite{UT84}).
Let us consider the pseudo-difference operators 
\beq\label{mkp1}
{\cal L}=e^{\eta \p_x}+\sum_{k\geq 0}U_k(x) e^{-k\eta \p_x}, \quad
\bar {\cal L}=c(x)e^{-\eta \p_x}+\sum_{k\geq 0}\bar U_k(x) e^{k\eta \p_x},
\eeq
where $e^{\eta \p_x}$ is the shift operator ($e^{\pm \eta \p_x}f(x)=f(x\pm \eta )$) and
the coefficient functions $U_k$, $\bar U_k$ 
are functions of $x$, ${\bf t}$ and $\bar {\bf t}$.
They are the Lax operators of the 2DTL hierarchy.
The equations of the hierarchy are differential-difference
equations for the functions $U_k$, $\bar U_k$. They are encoded in the Lax equations
\beq\label{mkp2}
\p_{t_m}{\cal L}=[{\cal A}_m, {\cal L}], \quad
\p_{t_m}\bar {\cal L}=[{\cal A}_m, \bar {\cal L}]
\qquad {\cal A}_m=({\cal L}^m)_{\geq 0},
\eeq
\beq\label{mkp2a}
\p_{\bar t_m}{\cal L}=[\bar {\cal A}_m, {\cal L}], \quad
\p_{\bar t_m}\bar {\cal L}=[\bar {\cal A}_m, \bar {\cal L}]
\qquad \bar {\cal A}_m=(\bar {\cal L}^m)_{< 0},
\eeq
where $\displaystyle{\Bigl (\sum_{\z} U_k e^{k\eta \p_x}\Bigr )_{\geq 0}=
\sum_{k\geq 0} U_k e^{k\eta \p_x}}$,
$\displaystyle{\Bigl (\sum_{k\in \z} U_k e^{k\eta \p_x}\Bigr )_{< 0}=
\sum_{k<0} U_k e^{k\eta \p_x}}$
For example, ${\cal A}_1=e^{\eta \p_x}+U_0(x)$, $\bar {\cal A}_1=c(x)e^{-\eta \p_x}$.

An equivalent formulation is through the zero
curvature (Zakharov-Shabat) equations
\beq\label{mkp3}
\p_{t_n}{\cal A}_m -\p_{t_m}{\cal A}_n +[{\cal A}_m, {\cal A}_n]=0,
\eeq
\beq\label{mkp3a}
\p_{\bar t_n}{\cal A}_m -\p_{t_m}\bar {\cal A}_n +[{\cal A}_m, \bar {\cal A}_n]=0,
\eeq
\beq\label{mkp3b}
\p_{\bar t_n}\bar {\cal A}_m -\p_{\bar t_m}\bar {\cal A}_n +[\bar {\cal A}_m, \bar {\cal A}_n]=0.
\eeq
In particular, at $n=1$, $m=2$ we obtain from (\ref{mkp3})
\beq\label{mkp4}
\left \{ \begin{array}{l}
\p_{t_1}\Bigl (U_0(x)+U_0(x-\eta )\Bigr )=U_1(x+\eta )-U_1(x-\eta )
\\ \\
\p_{t_2}U_0(x)=\p_{t_1}\Bigl (U_0^2(x)+U_1(x+\eta )+U_1(x)\Bigr ).
\end{array}
\right.
\eeq
Excluding $U_1$ from this system, one gets the mKP equation for
$v(x)=U_0(x)$:
\beq\label{mkp5}
\p_{t_2}\Bigl (v(x+\eta )-v(x)\Bigr )=
\p_{t_1}^2 \Bigl (v(x+\eta )+v(x)\Bigr ) +\p_{t_1}
\Bigl (v^2(x+\eta )-v^2(x)\Bigr ).
\eeq
From (\ref{mkp3a}) at $m=n=1$ we have
$$
\left \{ \begin{array}{l}
\p_{t_1}\log c(x)=v(x)-v(x-\eta )
\\ \\
\p_{\bar t_1} v(x)=c(x)-c(x+\eta ).
\end{array}
\right.
$$
Excluding $v(x)$, we get the second order differential-difference equation for 
$c(x)$:
$$
\p_{t_1}\p_{\bar t_1}\log c(x)=2c(x)-c(x+\eta )-c(x-\eta )
$$
which is one of the forms of the 2D Toda equation. After the change of variables
$c(x)=e^{\varphi (x)-\varphi (x-\eta )}$ it acquires the most familiar form
\beq\label{mkp5a}
\p_{t_1}\p_{\bar t_1}\varphi (x)=e^{\varphi (x)-\varphi (x-\eta )}-
e^{\varphi (x+\eta )-\varphi (x)}.
\eeq

The zero curvature equations
are compatibility conditions for the auxiliary linear problems
\beq\label{mkp6}
\p_{t_m}\psi ={\cal A}_m (x)\psi , \quad
\p_{\bar t_m}\psi =\bar {\cal A}_m (x)\psi ,
\eeq
where the wave function $\psi$ depends on a spectral parameter $z$: 
$\psi =\psi (z; {\bf t})$. The wave function has the following expansion in powers of
$z$:
\beq\label{mkp7}
\psi =z^{x/\eta}
e^{\xi ({\bf t}, z)}\left (1+ \frac{\xi_1(x, {\bf t}, \bar {\bf t})}{z}+
\frac{\xi_2(x, {\bf t}, \bar {\bf t})}{z^2}+\ldots \right ),
\eeq
where
\beq\label{mkp8}
\xi ({\bf t}, z)=\sum_{k\geq 1}t_k z^k.
\eeq
The wave operator is the pseudo-difference operator of the form
\beq\label{mkp9}
{\cal W}(x)=1+\xi_1(x)e^{-\eta \p_x}+\xi_2(x)e^{-2\eta \p_x}+\ldots \, 
\eeq
with the same coefficient functions $\xi_k$ as in 
(\ref{mkp7}), then the wave function can be written as
\beq\label{mkp10}
\psi = {\cal W}(x)z^{x/\eta} e^{\xi ({\bf t}, z)}.
\eeq
The adjoint wave function $\psi^{\dag}$ is defined by the formula
\beq\label{mkp11}
\psi^{\dag}=({\cal W}^{\dag}(x\! -\! \eta ))^{-1}z^{-x/\eta} e^{-\xi ({\bf t}, z)}
\eeq
(see, e.g., \cite{Z18}), 
where the adjoint difference operator is defined according to the rule
$(f(x) \circ e^{n\eta \p_x})^{\dag}=e^{-n\eta \p_x}\circ f(x)$. The auxiliary linear
problems for the adjoint wave function have the form
\beq\label{mkp12}
-\p_{t_m}\psi ^{\dag}={\cal A}_{m}^{\dag}(x\! -\! \eta )\psi^{\dag}.
\eeq
In particular, we have:
\beq\label{mkp13}
\begin{array}{l}
\phantom{-}\p_{t_1}\psi (x)=\psi (x+\eta )+v(x)\psi (x), 
\\ \\
-\p_{t_1}\psi^{\dag} (x)=\psi^{\dag} (x-\eta )+v(x-\eta )\psi^{\dag} (x),
\end{array}
\eeq
\beq\label{mkp13a}
\p_{\bar t_1}\psi (x)=c(x)\psi (x-\eta ).
\eeq

A common solution to the 2DTL hierarchy is provided by the tau-function $\tau =
\tau (x, {\bf t}, \bar {\bf t})$ \cite{DJKM83,JM83}. 
The hierarchy is encoded in the bilinear relation
\beq\label{mkp14b}
\begin{array}{c}
\displaystyle{\oint_{\infty}z^{\frac{x-x'}{\eta}-1}e^{\xi ({\bf t}, z)-\xi ({\bf t}', z)}
\tau \Bigl (x, {\bf t}-[z^{-1}], \bar {\bf t}\Bigr )
\tau \Bigl (x'+\eta , {\bf t}'+[z^{-1}], \bar {\bf t}'\Bigr )dz}
\\ \\
\displaystyle{=\, \oint_{0}z^{\frac{x-x'}{\eta}-1}
e^{\xi (\bar {\bf t}, z^{-1})-\xi (\bar {\bf t}', z^{-1})}
\tau \Bigl (x+\eta , {\bf t}, \bar {\bf t}-[z]\Bigr )
\tau \Bigl (x' , {\bf t}', \bar {\bf t}'+[z]\Bigr )dz
}
\end{array}
\eeq
valid for all $x, x'$, ${\bf t}, {\bf t}'$, $\bar {\bf t}, \bar {\bf t}'$, where
\beq\label{mkp15}
{\bf t}\pm [z]=\Bigl \{t_1\pm z, t_2\pm \frac{1}{2}z^{2}, 
t_3\pm \frac{1}{3}z^{3}, \ldots \Bigl \}.
\eeq
The integration contour in the left hand side 
is a big circle around infinity separating the singularities
coming from the exponential factor from those coming from the tau-functions. 
The integration contour in the right hand side 
is a small circle around zero separating the singularities
coming from the exponential factor from those coming from the tau-functions.
In particular, setting $x=x'$, $\bar {\bf t}=\bar {\bf t}'$, one obtains from 
(\ref{mkp14b}) the bilinear relation for the mKP hierarchy
\beq\label{mkp14}
\frac{1}{2\pi i}
\oint_{\infty}z^{-1}e^{\xi ({\bf t}, z)-\xi ({\bf t}', z)}
\tau \Bigl (x, {\bf t}-[z^{-1}]\Bigr )\tau \Bigl (x+\eta , {\bf t}'+[z^{-1}]\Bigr )dz =
\tau (x+\eta , {\bf t})\tau (x, {\bf t}').
\eeq
%\beq\label{mkp14a}
%\frac{1}{2\pi i}
%\oint_{\infty}ze^{\xi ({\bf t}, z)-\xi ({\bf t}', z)}
%\tau \Bigl (x+\eta , {\bf t}-[z^{-1}]\Bigr )\tau \Bigl (x, {\bf t}'+[z^{-1}]\Bigr )dz =0
%\eeq

Consequences of the bilinear
relations (which are in fact equivalent to the whole hierarchy, 
see \cite{TT95}) are the equations
\beq\label{mkp16}
\begin{array}{c}
\mu \tau (x+\eta , {\bf t}+[\lambda ^{-1}]-[\mu^{-1}], \bar {\bf t})
\tau (x, {\bf t}, \bar {\bf t})-
\lambda \tau (x+\eta , {\bf t}, \bar {\bf t})
\tau (x, {\bf t}+[\lambda ^{-1}]-[\mu^{-1}], \bar {\bf t})
\\ \\
+\, (\lambda -\mu )\tau (x+\eta , {\bf t}+[\lambda ^{-1}], \bar {\bf t})
\tau (x, {\bf t}-[\mu^{-1}], \bar {\bf t})=0,
\end{array}
\eeq
\beq\label{mkp16a}
\begin{array}{c}
\tau (x, {\bf t}-[\lambda^{-1}], \bar {\bf t})
\tau (x, {\bf t}, \bar {\bf t}-[\nu ])-
\tau (x, {\bf t}, \bar {\bf t})
\tau (x, {\bf t}-[\lambda^{-1}], \bar {\bf t}-[\nu ])
\\ \\
=\, \nu \lambda^{-1} \tau (x+\eta , {\bf t}, \bar {\bf t}-[\nu ])
\tau (x-\eta , {\bf t}-[\lambda^{-1}], \bar {\bf t}).
\end{array}
\eeq
There is also an equation similar to (\ref{mkp16}) with shifts of the negative times.
Together with the tau-function $\tau$ it is convenient to introduce another 
tau-function, $\tau'$, which differs from $\tau$ by a simple factor:
\beq\label{mkp16b}
\tau '(x, {\bf t}, \bar {\bf t})=\exp \left (\sum_{k\geq 1}kt_k \bar t_k\right )
\tau (x, {\bf t}, \bar {\bf t}).
\eeq

The coefficient functions of the Lax operators can be expressed through the 
tau-function. In particular,
\beq\label{mkp17}
U_0(x)=v(x)=\p_{t_1}\log \frac{\tau (x+\eta )}{\tau (x)}, \quad
c(x)=\frac{\tau (x+\eta )\tau (x-\eta )}{\tau^2(x)}.
\eeq
After this substitution the mKP equation (\ref{mkp5}) becomes
\beq\label{mkp18}
\p_{t_2}\log \frac{\tau (x+\eta )}{\tau (x)}=\p_{t_1}^2
\log (\tau (x+\eta )\tau (x))+\Bigl (\p_{t_1}\log \frac{\tau (x+\eta )}{\tau (x)}\Bigr )^2,
\eeq
which can be also represented in the bilinear form
$$
\tau (x) \p_{t_2}\tau (x+\eta )- \tau (x+\eta ) \p_{t_2}\tau (x)=
\tau (x+\eta )\p_{t_1}^2 \tau (x)-2\p_{t_1}\tau (x+\eta )\p_{t_1}\tau (x)+
\tau (x)\p_{t_1}^2 \tau (x+\eta ).
$$
The Toda equation (\ref{mkp5a}) becomes
\beq\label{mkp18a}
\p_{t_1}\p_{\bar t_1}\log \tau (x)=1-\frac{\tau (x+\eta )\tau (x-\eta )}{\tau^2(x)}.
\eeq

The wave function and its adjoint are expressed through the tau-function according to
the formulas
\beq\label{mkp19}
\psi =z^{x/\eta}e^{\xi ({\bf t}, z)}\frac{\tau (x, {\bf t}-[z^{-1}], 
\bar {\bf t})}{\tau (x, {\bf t}, \bar {\bf t})},
\eeq
\beq\label{mkp20}
\psi^{\dag} =z^{-x/\eta}e^{-\xi ({\bf t}, z)}
\frac{\tau (x, {\bf t}+[z^{-1}], \bar {\bf t})}{\tau (x, {\bf t}, \bar {\bf t})}.
\eeq
One may also introduce the complimentary wave functions $\bar \psi$, $\bar \psi^{\dag}$ 
by the formulas
\beq\label{mkp19a}
\bar \psi =z^{x/\eta}e^{\xi (\bar {\bf t}, z^{-1})}\frac{\tau (x+\eta , {\bf t}, 
\bar {\bf t}-[z])}{\tau (x, {\bf t}, \bar {\bf t})},
\eeq
\beq\label{mkp20a}
\bar \psi^{\dag} =z^{-x/\eta}e^{-\xi (\bar {\bf t}, z^{-1})}
\frac{\tau (x-\eta , {\bf t}, \bar {\bf t}+[z])}{\tau (x, {\bf t}, \bar {\bf t})}.
\eeq
They satisfy the same auxiliary linear problems as the wave functions $\psi$, $\psi^{\dag}$.
It will be more convenient for us to work with the renormalized wave functions
\beq\label{mkp19b}
\phi (x)=\frac{\tau (x)}{\tau (x+\eta )}\, \bar \psi (x)=
z^{x/\eta}e^{\xi (\bar {\bf t}, z^{-1})}\frac{\tau (x+\eta , {\bf t}, 
\bar {\bf t}-[z])}{\tau (x+\eta , {\bf t}, \bar {\bf t})},
\eeq
\beq\label{mkp20b}
\phi^{\dag}(x)=\frac{\tau (x)}{\tau (x-\eta )}\, \bar \psi^{\dag}(x)=
z^{-x/\eta}e^{-\xi (\bar {\bf t}, z^{-1})}
\frac{\tau (x-\eta , {\bf t}, \bar {\bf t}+[z])}{\tau (x-\eta , {\bf t}, \bar {\bf t})}.
\eeq
It is easy to check that they satisfy the linear equations
\beq\label{mkp20c}
\p_{\bar t_1}\phi (x)=\phi (x-\eta )-\bar v(x)\phi (x), \quad
-\p_{\bar t_1}\phi^{\dag} (x)=\phi^{\dag} (x+\eta )-\bar v(x-\eta )\phi ^{\dag} (x),
\eeq
where $\displaystyle{\bar v(x)=\p_{\bar t_1}\log \frac{\tau (x+\eta )}{\tau (x}}$.

Finally, let us point out useful corollaries of the bilinear relation (\ref{mkp14}).
Differentiating it with respect to $t_m$ and putting ${\bf t}={\bf t}'$ after that, we obtain:
\beq\label{mkp21}
\begin{array}{c}
\displaystyle{
\frac{1}{2\pi i}
\oint_{\infty}z^{m-1}
\tau \Bigl (x, {\bf t}-[z^{-1}]\Bigr )\tau \Bigl (x+\eta , {\bf t}+[z^{-1}]\Bigr )dz}
\\ \\
\displaystyle{ =
\p_{t_m}\tau (x+\eta , {\bf t})\tau (x, {\bf t})-
\p_{t_m}\tau (x, {\bf t})\tau (x+\eta , {\bf t})}
\end{array}
\eeq
or
\beq\label{mkp22}
\res_{\infty}\, \Bigl (z^m \psi (x)\psi^{\dag}(x+\eta )\Bigr )=\p_{t_m}
\log \frac{\tau (x+\eta )}{\tau (x)}.
\eeq
In a similar way, differentiating the bilinear relation (\ref{mkp14b}) 
with respect to $\bar t_m$ and putting $x=x'$, ${\bf t}={\bf t}'$, 
$\bar {\bf t}=\bar {\bf t}'$ after that,
we obtain the relation
\beq\label{mkp22a}
\res_{0}\, \Bigl (z^{-m} \phi (x)\phi^{\dag}(x+\eta )\Bigr )=-\p_{\bar t_m}
\log \frac{\tau (x+\eta )}{\tau (x)}.
\eeq
Here $\displaystyle{\res_{\infty}}$, $\displaystyle{\res_{0}}$ are defined according to the convention $\displaystyle{\res_{\infty}}(z^{-n})=\delta_{n1}$, 
$\displaystyle{\res_{0}}(z^{-n})=\delta_{n1}$.

\section{Trigonometric solutions to the mKP equation}

We are going to consider solutions which are trigonometric (i.e. single-periodic)
functions of the variable $x$. 
For trigonometric solutions the tau-function has the form
\beq\label{ts1}
\tau (x, {\bf t})=\prod_{i=1}^{N}\Bigl ( e^{2\gamma x}-e^{2\gamma x_i({\bf t})}\Bigr ).
\eeq
(In this section we ignore the dependence on the 
negative times keeping them fixed to zero.) 
This function has a single period $\pi i/\gamma$ in the complex plane. As in \cite{Z19a},
we pass to the exponentiated variables
\beq\label{ts2}
w=e^{2\gamma x}, \quad w_i=e^{2\gamma x_i}.
\eeq
In these variables, the tau-function becomes a polynomial of degree $N$ 
with roots $w_i$ which
are supposed to be distinct: $\displaystyle{\tau =\prod_i (w-w_i)}$. The function
$v(x)=\p_{t_1}\log \Bigl (\tau (x+\eta )/\tau (x) \Bigr )$ is
\beq\label{ts3}
v(x)=\sum_i \left ( \frac{\dot w_i}{w-w_i}-\frac{\dot w_i}{qw-w_i}\right ),
\eeq
where
\beq
q=e^{2\gamma \eta}
\eeq
and here and below dot means the $t_1$-derivative. 

We begin with the investigation of the $t_1$-dynamics of the poles. 
The ansatz for the $\psi$-function depending on the spectral parameter $z$ 
suggested by 
equation (\ref{mkp19}) is
\beq\label{ts4}
\psi =z^{x/\eta} e^{t_1z}\left (1+\sum_i \frac{2\gamma c_i}{w-w_i}\right ),
\eeq
where we have put $t_k=0$ for $k\geq 2$. The coefficients $c_i$ depend on 
${\bf t}$ and on $z$. Substituting $\psi$ and $v$ into the first auxiliary linear
problem in (\ref{mkp13}) $-\p_{t_1} \psi (x)+\psi (x+\eta )+v(x)\psi (x)=0$, we get:
$$
-z\sum_i \frac{c_i}{w-w_i}-\sum_i \frac{\dot c_i}{w-w_i}-
\sum_i \frac{\dot w_ic_i}{(w-w_i)^2}+\sum_i \frac{q^{-1}c_i}{w-q^{-1}w_i}
$$
$$
+\frac{1}{2\gamma}\sum_i \left (\frac{\dot w_i}{w-w_i}-
\frac{\dot w_i q^{-1}}{w-q^{-1}w_i}\right  )+
\sum_i \left (\frac{\dot w_i}{w-w_i}-
\frac{\dot w_i q^{-1}}{w-q^{-1}w_i}\right  )\sum_k \frac{c_k}{w-w_k}=0.
$$
The left hand side is a rational function of $w$ vanishing at infinity
with simple poles at $w=w_i$ and $w=q^{-1}w_i$ (the second order poles
cancel identically). We should equate residues at the poles to zero. 
This gives the following system of linear equations for the coefficients $c_i$:
\beq\label{ts5}
\left \{ \begin{array}{l}
\displaystyle{zc_i -q \sum_k \frac{\dot w_i c_k}{w_i-qw_k}=\frac{1}{2\gamma}\, \dot w_i}
\\ \\
\displaystyle{\dot c_i=c_i \left (\sum_{k\neq i}\frac{\dot w_k}{w_i-w_k}
-\sum_k \frac{\dot w_k}{qw_i-w_k}\right )+
\sum_{k\neq i}\frac{\dot w_i c_k}{w_i-w_k}-
q \sum_k \frac{\dot w_i c_k}{w_i-qw_k}}.
\end{array}
\right.
\eeq
In a similar way, the adjoint linear problem
$\p_{t_1} \psi^\dag (x)+\psi^\dag (x-\eta )+v(x-\eta )\psi^{\dag} (x)=0$
with the ansatz for the $\psi^{\dag}$-function
\beq\label{ts6}
\psi^{\dag} =z^{-x/\eta} e^{-t_1z}\left (1+\sum_i \frac{2\gamma c_i^{*}}{w-w_i}\right )
\eeq
leads to the linear equations for the coefficients $c_i^*$:
\beq\label{ts7}
\left \{ \begin{array}{l}
\displaystyle{zc_i^* - \sum_k \frac{\dot w_i c_k^*}{w_k-qw_i}=-\, 
\frac{1}{2\gamma}\, \dot w_i}
\\ \\
\displaystyle{\dot c_i^*=c_i^* \left (\sum_{k\neq i}\frac{\dot w_k}{w_i-w_k}
+\sum_k \frac{q\dot w_k}{w_i-qw_k}\right )+
\sum_{k\neq i}\frac{\dot w_i c_k^*}{w_i-w_k}-
\sum_k \frac{\dot w_i c_k^*}{qw_i-w_k}}.
\end{array}
\right.
\eeq
After the gauge transformation $\tilde c_i =c_i w_i^{-1/2}$, 
$\tilde c^*_i =c^*_i w_i^{-1/2}$ the above conditions 
can be written in the matrix form
\beq\label{ts8}
\Bigl (zI-q^{1/2}L\Bigr )\tilde {\bf c}=\dot X W^{1/2}{\bf e}, \quad
\p_{t_1}\tilde {\bf c}=M\tilde {\bf c},
\eeq
\beq\label{ts9}
\tilde {\bf c}^* \dot X^{-1}
\Bigl (zI-q^{-1/2}L\Bigr )=-{\bf e}^TW^{1/2}, \quad
\p_{t_1}\tilde {\bf c}^*=-\tilde {\bf c}^*\tilde M,
\eeq
where $\tilde {\bf c}=(\tilde c_1, \ldots , \tilde c_N)^T$ is a column vector,
$\tilde {\bf c}^*=(\tilde c^*_1, \ldots , \tilde c^*_N)$ is a row vector,
${\bf e}=(1, 1, \ldots , 1)^T$, $I$ is the identity matrix 
and the matrices $X$, $W$, $L$, $M$, $\tilde M$ are
\beq\label{ts10}
X=\mbox{diag}\, (x_1, x_2, \ldots , x_N), \quad 
W=\mbox{diag}\, (w_1, w_2, \ldots , w_N),
\eeq
\beq\label{ts11}
L_{ij}=2\gamma q^{1/2}\, \frac{\dot x_i w_i^{1/2}w_j^{1/2}}{w_i-qw_j},
\eeq
\beq\label{ts12}
M_{ij}=\gamma \delta_{ij}\left (\sum_{k\neq i}\frac{w_i\! +\! w_k}{w_i\! -\! w_k}\, \dot x_k
\! -\! \sum_k \frac{qw_i\! +\! w_k}{qw_i\! -\! w_k}\, \dot x_k \right )
+2\gamma \frac{\dot x_i w_i^{1/2}w_j^{1/2}}{w_i-w_j}\, (1\! -\! \delta_{ij})
-2\gamma q \frac{\dot x_i w_i^{1/2}w_j^{1/2}}{w_i-qw_j},
\eeq
\beq\label{ts13}
\tilde 
M_{ji}=-\gamma \delta_{ij}\left (\sum_{k\neq i}\frac{w_i\! +\! w_k}{w_i\! -\! w_k}\, \dot x_k
\! -\! \sum_k \frac{w_i\! +\! qw_k}{w_i\! -\! qw_k}\, \dot x_k \right )
+2\gamma \frac{\dot x_i w_i^{1/2}w_j^{1/2}}{w_j-w_i}\, (1\! -\! \delta_{ij})-
2\gamma \frac{\dot x_i w_i^{1/2}w_j^{1/2}}{w_j-qw_i}.
\eeq
The following commutation relation can be checked directly:
\beq\label{ts14}
q^{-1/2}WL-q^{1/2}LW =W^{-1/2}\dot W E W^{1/2}.
\eeq
Here $E={\bf e}\otimes {\bf e}^T$ is the rank 1 matrix with matrix elements 
$E_{ij}=1$. This commutation relation will be important in what follows. 

The system of linear equations (\ref{ts8}) is overdetermined. 
Taking the $t_1$-derivative of the first equation in (\ref{ts8})
and substituting the second equation, one obtains
the compatibility condition of the system in the form
\beq\label{ts15}
\Bigl (\dot L+[L,M]\Bigr )\tilde {\bf c}+q^{-1/2}\Bigl (
\ddot X +\gamma \dot X^2 -M\dot X\Bigr )W^{1/2}{\bf e}=0.
\eeq
A straightforward calculation shows that
$$
\dot L+[L,M]=RL,
$$
$$
\Bigl (\ddot X +\gamma \dot X^2 -M\dot X\Bigr )W^{1/2}{\bf e}=
R\dot X W^{1/2}{\bf e},
$$
where
$$
R=\ddot X \dot X^{-1} +D^++D^- -2D^0
$$
and the diagonal matrices $D^{\pm}$, $D_0$ are
$$
D^{\pm}_{ij}=\delta_{ij}\gamma 
\sum_{k\neq i}\frac{q^{\pm 1}w_i+w_k}{q^{\pm 1}w_i-w_k}\, \dot x_k, \quad
D^{0}_{ij}=\delta_{ij}\gamma 
\sum_{k\neq i}\frac{w_i+w_k}{w_i-w_k}\, \dot x_k.
$$
Therefore, the compatibility condition takes the form $R\tilde {\bf c}=0$ which means that
$R_{ii}=0$ for all $i$. This leads to the equations of motion of the 
trigonometric Ruijsenaars-Schneider
model
\beq\label{ts16}
\begin{array}{lll}
\ddot x_i &=& \displaystyle{-\gamma \sum_{k\neq i}\dot x_i \dot x_k \Bigl (
\coth (\gamma (x_{ik}+\eta ))+\coth (\gamma (x_{ik}-\eta ))-2\coth (\gamma x_{ik})\Bigr )}
\\ && \\
&=&\displaystyle{\sum_{k\neq i}\dot x_i \dot x_k\frac{2\gamma \sinh^2(\gamma \eta )
\cosh (\gamma x_{ik})}{\sinh (\gamma x_{ik})
\sinh (\gamma (x_{ik}+\eta ))\sinh (\gamma (x_{ik}-\eta ))},}
\end{array}
\eeq
where $x_{ik}=x_i-x_k$. The matrix equation $\dot L+[L,M']=0$ with
\beq\label{ts16a}
L_{ij}= \frac{\gamma \dot x_i}{\sinh (\gamma (x_{ij}-\eta ))},
\eeq
\beq\label{ts16b}
M_{ij}'=\gamma \delta_{ij}\left (\sum_{k\neq i}\dot x_k \coth (\gamma x_{ik})-
\sum_{k}\dot x_k \coth (\gamma (x_{ik}+\eta ))\right )+(1-\delta_{ij})\,
\frac{\gamma \dot x_i}{\sinh (\gamma x_{ij})}
\eeq
provides the Lax representation for them.

Equations (\ref{ts16}) are Hamiltonian with the Hamiltonian
\beq\label{ts17}
H_1=\sum_i e^{\eta p_i}\prod_{k\neq i}
\frac{\sinh (\gamma (x_{ik}+\eta ))}{\sinh (\gamma x_{ik})}
\eeq
and the canonical Poisson brackets between $p_i$ and $x_i$. The velocity is given by
$$
\dot x_i=\frac{\p H_1}{\p p_i}=\eta e^{\eta p_i}\prod_{k\neq i}
\frac{\sinh (\gamma (x_{ik}+\eta ))}{\sinh (\gamma x_{ik})}.
$$
The Lax representation implies that the higher conserved quantities are 
$\mbox{tr}\, L^m$. It is proved in \cite{Ruij87} that they are in involution, 
i.e., the system is integrable. 

Let us consider the transformation of the phase space coordinates
$(p_i, x_i)\to (u_i , x_i')$, where $x_i'=x_i$ and
\beq\label{ts18a}
u_i=\log \dot x_i = \eta p_i +\sum_{k\neq i} \log 
\frac{\sinh (\gamma (x_{ik}+\eta ))}{\sinh (\gamma x_{ik})} +\log \eta .
\eeq
Then the derivatives transform as follows:
\beq\label{ts18}
\frac{\p f}{\p p_i}=\sum_k \frac{\p u_k}{\p p_i}\, \frac{\p f}{\p u_k}+
\sum_k \frac{\p x_k'}{\p p_i}\, \frac{\p f}{\p x_k'}=\eta \frac{\p f}{\p u_i},
\eeq
\beq\label{ts19}
\begin{array}{lll}
\displaystyle{\frac{\p f}{\p x_i}}&=&\displaystyle{
\sum_k \frac{\p u_k}{\p x_i}\, \frac{\p f}{\p u_k}+
\sum_k \frac{\p x_k'}{\p x_i}\, \frac{\p f}{\p x_k'}}
\\ && \\
&=&\displaystyle{\frac{\p f}{\p x_i'}+\gamma \frac{\p f}{\p u_i}
\sum_{l\neq i} (\coth (\gamma (x_{il}+\eta ))-\coth (\gamma x_{il}))}
\\ &&\\
&&\phantom{aaaa} +\, \displaystyle{\gamma \sum_{k\neq i}\frac{\p f}{\p u_k}
(\coth (\gamma (x_{ik}+\eta ))-\coth (\gamma x_{ik})).}
\end{array}
\eeq
At this point we finish the discussion of the $t_1$-dynamics of poles and pass to
the higher times in the next section. 

\section{The higher Hamiltonian equations} 

\subsection{Positive times}

In order to study the dynamics of poles in the higher positive times ${\bf t}$, we use 
the relation (\ref{mkp22}), which, after the substitution of the wave functions
for the trigonometric solutions, takes the form
$$
\frac{1}{2\pi i}\oint_{\infty}z^{m-1}\left (1+\sum_i 
\frac{2\gamma c_i}{w-w_i}\right )\left (1+\sum_k 
\frac{2\gamma c_k^*}{qw-w_k}\right )dz=
\sum_i \left (\frac{\p_{t_m}w_i}{w-w_i}-\frac{\p_{t_m}w_i}{qw-w_i}\right ).
$$
The both sides are rational functions of $w$ with simple poles at $w=w_i$ and
$w=q^{-1}w_i$ vanishing at infinity. Identifying the residues at the poles at
$w=w_i$ in the both sides, we obtain:
\beq\label{h1}
\p_{t_m}x_i=-2\gamma \res_{\infty} \, \Bigl (z^m \tilde c_i^* \dot w_i^{-1}
\tilde c_i \Bigr ).
\eeq
From (\ref{ts8}), (\ref{ts9}) we have:
$$
\tilde {\bf c}=\frac{1}{2\gamma}(zI-q^{1/2}L)^{-1}\dot W W^{-1/2}{\bf e},
\quad
\tilde {\bf c}^*=-\frac{1}{2\gamma}\, {\bf e}^T W^{1/2}(zI-q^{-1/2}L)^{-1}\dot W W^{-1}.
$$
Substituting this into (\ref{h1}), we get:
$$
\p_{t_m}x_i=\frac{1}{2\gamma}\res_{\infty}\sum_{k,k'}
\left [ z^m w_k^{1/2}\Bigl (\frac{1}{zI-q^{-1/2}L}\Bigr )_{ki}w_i^{-1}
\Bigl (\frac{1}{zI-q^{1/2}L}\Bigr )_{ik'}\dot w_{k'}w_{k'}^{-1/2}\right ]
$$
$$
=\frac{1}{2\gamma}\res_{\infty} \, \mbox{tr}\left (
z^m \dot W W^{-1/2} EW^{1/2}\frac{1}{zI-q^{-1/2}L} E_i W^{-1}\frac{1}{zI-q^{1/2}L}\right ),
$$
where $E_i$ is the diagonal matrix with the matrix elements $(E_i)_{jk}=\delta_{ij}\delta_{ik}$. 
Using the commutation relation (\ref{ts14}), we have:
$$
\p_{t_m}x_i=\frac{1}{2\gamma}\res_{\infty} \, \mbox{tr}\left (
z^m (q^{-1/2}WL-q^{1/2}LW) \frac{1}{zI-q^{-1/2}L} \, E_i W^{-1}\frac{1}{zI-q^{1/2}L}\right )
$$
$$
=\frac{1}{2\gamma}\res_{\infty} \, \mbox{tr}\left (z^m \Bigl (
E_i\, \frac{1}{zI-q^{-1/2}L}-E_i\, \frac{1}{zI-q^{1/2}L}\Bigr )\right ).
$$
Next, we use the easily proved identity
$$
E_iL=\dot x_i \frac{\p L}{\p \dot x_i}=\frac{\p L}{\p u_i}=\eta^{-1}\frac{\p L}{\p p_i}
$$
(see (\ref{ts18})) to continue the chain of equalities:
$$
\p_{t_m}x_i=\frac{1}{2\gamma \eta}\res_{\infty} \, \mbox{tr}\left (z^m
\left (\frac{\p L}{\p p_i}\, \frac{L^{-1}}{zI-q^{-1/2}L}-
\frac{\p L}{\p p_i}\, \frac{L^{-1}}{zI-q^{1/2}L}\right ) \right )
$$
\beq\label{h2}
=\frac{1}{2\gamma \eta}\, 
(q^{-m/2}-q^{m/2})\, \mbox{tr}  \left (\frac{\p L}{\p p_i}\, L^{m-1}\right )=
-\frac{\sinh (m\gamma \eta )}{m\gamma \eta }\frac{\p}{\p p_i}\, \mbox{tr}\, L^m.
\eeq
In this way we have obtained one half of the Hamiltonian equations for the higher flows
\beq\label{h3}
\p_{t_m}x_i=\frac{\p H_m}{\p p_i}, \quad
H_m=-\frac{\sinh (m\gamma \eta )}{m\gamma \eta }\, \mbox{tr}\, L^m,
\eeq
where the Lax matrix $L$ is given by (\ref{i2}). In particular, the Hamiltonian 
$H_1$ coincides with (\ref{ts17}). 

The derivation of the second half of the Hamiltonian equations is more involved.
The idea of the derivation is the same as in \cite{Iliev07}.
First of all, we note that (\ref{h3}) can be written in the form
$$
\p_{t_m}x_i=-m\eta \kappa_m \mbox{tr}\, (E_i L^m), \quad 
\kappa_m= \frac{\sinh (m\gamma \eta )}{m\gamma \eta }.
$$
Differentiating this equality with respect to $t_1$ and using the Lax equation,
we get:
$$
\p_{t_m}\dot x_i=-m\eta \kappa_m \mbox{tr}\, (E_i[M', L^m])=
-m\eta \kappa_m \mbox{tr}\, (L^m [E_i, M']).
$$
Now we apply $\p_{t_m}$ to equation (\ref{ts18a}):
$$
\p_{t_m}p_i=\eta^{-1}\p_{t_m}\log \dot x_i -\eta^{-1}\sum_j\sum_{l\neq i}
\frac{\p}{\p x_j}\log \frac{\sinh (\gamma (x_{il}+\eta ))}{\sinh (\gamma x_{il})}\,
\p_{t_m} x_j
$$
$$
=-m \kappa_m \dot x_i^{-1}\mbox{tr}\, (L^m [E_i, M'])+m\kappa_m
\sum_j\sum_{l\neq i}
\frac{\p}{\p x_j}\log \frac{\sinh (\gamma (x_{il}+\eta ))}{\sinh (\gamma x_{il})}\,
\mbox{tr}\, (E_j L^m)
$$
$$
=-m\kappa_m \, \mbox{tr} \, \Bigl (A^{(i)}L^{m-1}\Bigr ),
$$
where the matrix $A^{(i)}$ is
\beq\label{h4}
A^{(i)}=\dot x_i^{-1}(LE_iM'-M'E_iL)-
\sum_j\sum_{l\neq i}
\frac{\p}{\p x_j}\log \frac{\sinh (\gamma (x_{il}+\eta ))}{\sinh (\gamma x_{il})}
\, E_j L.
\eeq
Note that the diagonal part of the matrix $M'$ (\ref{ts16b}) does not contribute, so  
instead of the matrix $M'$ here we can substitute its off-diagonal part
$$
A_{ij}=2\gamma (1-\delta_{ij})\, \frac{\dot x_i w_i^{1/2}w_j^{1/2}}{w_i-w_j}.
$$
Let us calculate the matrix elements:
$$
(LE_iA)_{jk}=\gamma \dot x_i L_{jk}\left (\frac{w_i+w_k}{w_i-w_k}-
\frac{qw_i+w_k}{qw_i-w_k}\right )(1-\delta_{ik}),
$$
$$
(AE_iL)_{jk}=-\gamma \dot x_i L_{jk}\left (\frac{w_i+w_j}{w_i-w_j}-
\frac{w_i+qw_k}{w_i-qw_k}\right )(1-\delta_{ij}),
$$
$$
\sum_l\sum_{r\neq i}
\frac{\p}{\p x_l}\log \frac{\sinh (\gamma (x_{ir}+\eta ))}{\sinh (\gamma x_{ir})}
\, (E_l L)_{jk}
$$
$$
=\gamma \delta_{ij}L_{jk}\sum_{r\neq i}\left (\frac{qw_i+w_r}{qw_i-w_r}-
\frac{w_i+w_r}{w_i-w_r}\right )-\gamma (1-\delta_{ij})L_{jk}
\left (\frac{qw_i+w_j}{qw_i-w_j}-
\frac{w_i+w_j}{w_i-w_j}\right ).
$$
Combining everything together, we obtain the matrix elements of the matrix
$A^{(i)}$:
\beq\label{h5}
\begin{array}{lll}
A^{(i)}_{jk}&=&\displaystyle{\gamma L_{jk}\left (\frac{w_i+w_k}{w_i-w_k}(1-\delta_{ik})-
\frac{w_i+qw_k}{w_i-qw_k}(1-\delta_{ij})+\frac{qw_i+w_j}{qw_i-w_j}(\delta_{ik}-\delta_{ij})
\right.}
\\ &&\\
&&\phantom{aa}\displaystyle{\left. -\delta_{ij}\sum_{r\neq i}\left (
\frac{qw_i+w_r}{qw_i-w_r}-\frac{w_i+w_r}{w_i-w_r}\right ) \right )}.
\end{array}
\eeq
Our next goal is to prove that 
\beq\label{h6a}
A^{(i)}=-\frac{\p L}{\p x_i}-[C^{(i)}, L],
\eeq
where the matrix $C^{(i)}$ is given by
\beq\label{h6}
C^{(i)}=\gamma \sum_{l}\frac{qw_l+w_i}{qw_l-w_i}\, E_l -\gamma \sum_{l\neq i}
\frac{w_l+w_i}{w_l-w_i}\, E_l.
\eeq
From this one concludes that
$$
\p_{t_m}p_i=-m\kappa_m \, \mbox{tr} \, \Bigl (A^{(i)}L^{m-1}\Bigr )=
m \kappa_m \, \mbox{tr}\left (\frac{\p L}{\p x_i}\, L^{m-1}\right )=
\kappa_m \frac{\p}{\p x_i}\, \mbox{tr}\, L^m.
$$
This yields the second half of the Hamiltonian equations for the higher flows:
\beq\label{h7}
\p_{t_m}p_i=-\frac{\p H_m}{\p x_i}.
\eeq

In order to prove the identity (\ref{h6a}), we calculate matrix elements of the 
right hand side and compare them with (\ref{h5}). Indeed, we have:
$$
\frac{\p L_{jk}}{\p x_i}=\gamma L_{jk}\left (\frac{w_j+qw_k}{w_j-qw_k}\,
(\delta_{ik}-\delta_{ij})+\frac{w_i+qw_j}{w_i-qw_j}\, (1-\delta_{ij})-
\frac{w_i+w_j}{w_i-w_j}\, (1-\delta_{ij})\right.
$$
$$
\left. +\, \delta_{ij}\sum_{r\neq i}\left (\frac{qw_i+w_r}{qw_i-w_r}-
\frac{w_i+w_r}{w_i-w_r}\right )\right ),
$$
$$
[C^{(i)}, L]_{jk}=\gamma L_{jk}\left (\frac{qw_j+w_i}{qw_j-w_i}-
\frac{qw_k+w_i}{qw_k-w_i}-\frac{w_j+w_i}{w_j-w_i} (1-\delta_{ij})+
\frac{w_k+w_i}{w_k-w_i} (1-\delta_{ik})\right ),
$$
and one can check that $A^{(i)}+\p L/\p x_i+[C^{(i)}, L]=0$. 

\subsection{Negative times}

In order to investigate the dynamics of zeros of the tau-function in the negative times,
we first consider the $\bar t_1$-evolution. We will work with the complimentary wave functions
$\phi$, $\phi^{\dag}$ given by (\ref{mkp19b}), (\ref{mkp20b}) for which we use the ansatz
\beq\label{n1}
\phi (x)=z^{x/\eta}e^{\bar t_1 z^{-1}}\left (1+\sum_i \frac{2\gamma b_i}{qw-w_i}\right ),
\eeq
\beq\label{n1a}
\phi^{\dag} (x)=z^{-x/\eta}e^{-\bar t_1 z^{-1}}\left (1+\sum_i 
\frac{2\gamma b_i^*}{q^{-1}w-w_i}\right ),
\eeq
where $b_i$, $b_i^*$ are some unknown coefficients depending on the times and on $z$ but not
on $x$. Substituting them into the linear equations (\ref{mkp20c}), we can write down the 
conditions of cancellation of the poles in the way similar to that of section 3. 
In fact the equations for $b_i$ are the same as for $c_i^{*}$ up to changing 
$z$ to $-z^{-1}$, $w$ to $qw$ and $\p_{t_1}$ to $\p_{\bar t_1}$. The equations
for $b_i^{*}$ and $c_i$ are connected in a similar way. 
Passing to $\tilde b_i=w_{i}^{-1/2}b_i$, $\tilde b_i^*=w_{i}^{-1/2}b_i^*$, we have,
after some algebra, in the notation of section 3:
\beq\label{n2}
\tilde {\bf b}^T (\p_{\bar t_1}X)^{-1} (z^{-1}I+q^{-1/2}\bar L)={\bf e}^T W^{1/2},
\eeq
\beq\label{n2a}
(z^{-1}I+q^{1/2}\bar L)\tilde {\bf b}^{*} =-\p_{\bar t_1}X\, W^{1/2}{\bf e},
\eeq
where $\tilde {\bf b}^T =(\tilde b_1, \ldots , \tilde b_N)$,
$\tilde {\bf b}^{*} =(\tilde b_1^*, \ldots , \tilde b_N^*)$ and the matrix $\bar L$ reads
\beq\label{n3}
\bar L_{ij}=2\gamma q^{1/2} \frac{\p_{\bar t_1}x_i\, w_i^{1/2}w_j^{1/2}}{w_i-qw_j}.
\eeq
This matrix satisfies the commutation relation (\ref{ts14}) with $\p_{\bar t_1}W$
instead of $\dot W =\p_{t_1}W$:
\beq\label{n4}
q^{-1/2}W\bar L-q^{1/2}\bar LW =W^{-1/2} \p_{\bar t_1} W \, E W^{1/2}.
\eeq
Using the relation (\ref{mkp22a}), we find, similarly to (\ref{h1}):
\beq\label{n5}
\p_{\bar t_m}x_i=-2\gamma \res_0 \Bigl (z^{-m-2}\tilde b_i^* (\p_{\bar t_1}w_i)^{-1}
\tilde b_i \Bigr ).
\eeq
Substituting here the solutions of linear systems (\ref{n2}), (\ref{n2a}) and using
(\ref{n4}), one can repeat the calculation from section 4.1 with the result
\beq\label{n6}
\p_{\bar t_m}x_i =(-1)^m \, \frac{\sinh (m\gamma \eta )}{\gamma} \, \mbox{tr}\,
(E_i \bar L^m).
\eeq

Our next goal is to derive a relation between the Lax matrices $L$ and $\bar L$. 
For this, we need a relation between the velocities $\dot x_i=\p_{t_1}x_i$ and 
$\p_{\bar t_1}x_i$. The latter can be derived from the Toda equation (\ref{mkp18a}).
Substituting the tau-function in the form $\tau =\prod_i (w-w_i)$, we get
$$
-\sum_i \frac{\p_{t_1}\p_{\bar t_1}w_i}{w-w_i}-\sum_i 
\frac{\p_{t_1}w_i \, \p_{\bar t_1}w_i}{(w-w_i)^2}=1-\prod_k 
\frac{(qw-w_k)(q^{-1}w-w_k)}{(w-w_k)^2}.
$$
Identifying the second order poles in the both sides, we obtain the relation
\beq\label{n7}
\p_{t_1}w_i \, \p_{\bar t_1}w_i=\frac{\prod_k(qw_i-w_k)(q^{-1}w_i-w_k)}{\prod_{l\neq i}
(w_i-w_l)^2}
\eeq
or
\beq\label{n8}
\p_{t_1}X \, \p_{\bar t_1}X =\frac{1}{4\gamma^2}\, W^{-2}U_+U_-,
\eeq
where the diagonal matrices $U_{\pm}$ are
\beq\label{n9}
(U_{\pm})_{ij}=\delta_{ij}\, \frac{\prod_k (w_i -q^{\pm 1}w_k)}{\prod_{l\neq i}(w_i-w_l)}.
\eeq
Next, we need the formula for the inverse of the Cauchy matrix
\beq\label{n10}
C_{ij}=\frac{1}{w_i-qw_j}.
\eeq
We have:
\beq\label{n11}
C^{-1}_{ij}=\frac{1}{qw_i-w_j}\, \frac{\prod_k (qw_i-w_k)(w_j-qw_k)}{q^{N-1}
\prod_{l\neq j} (w_j-w_l)\prod_{l'\neq i}(w_i-w_{l'})}
\eeq
or
\beq\label{n12}
C^{-1}=-qU_- C^T U_+.
\eeq
Now we write $L=2\gamma q^{1/2} \p_{t_1}X \, W^{1/2} C W^{1/2}$ and find, using
(\ref{n12}), (\ref{n8}):
$$
L^{-1}=\frac{q^{-1/2}}{2\gamma}W^{-1/2}C^{-1}W^{-1/2}(\p_{t_1}X)^{-1}
$$
$$
=-2\gamma q^{1/2} W^{-1/2}U_- C^T W^{-1/2} \p_{\bar t_1}X\, W^2 U_{-}^{-1}
$$
$$
=-2\gamma q^{1/2}W^{-1}U_- \Bigl (\p_{\bar t_1}X\, W^{1/2}C W^{1/2}\Bigr )^T 
(W^{-1}U_-)^{-1}
$$
$$
=-W^{-1}U_- \,  \bar L^T (W^{-1}U_-)^{-1}.
$$
We see that the matrix $-\bar L^T$ is connected with $L^{-1}$ by a similarity 
transformation with a diagonal matrix. Using the fact that $E_i \bar L=-\eta^{-1}
\p \bar L/\p p_i$, we can therefore rewrite equation (\ref{n6}) as
\beq\label{n13}
\p_{\bar t_m}x_i =- \frac{\sinh (m\gamma \eta )}{m\gamma \eta} \, 
\frac{\p}{\p p_i}\, \mbox{tr}\,
L^{-m} =\frac{\p \bar H_m}{\p p_i}
\eeq
which is one half of the Hamiltonian equations for the negative time flows. 

The derivation of the second half is straightforward. We note that
$$
\p_{\bar t_m}x_i =m\eta \kappa_m \, \mbox{tr}\, (E_iL^{-m}),
$$
where we use the notation of section 4.1. In the complete analogy with the calculation
in the previous subsection, we have
$$
\p_{\bar t_m}p_i=m\kappa_m \, \mbox{tr}\, (A^{(i)}L^{-m-1})
$$
with the same matrix $A^{(i)}$ (\ref{h4}). By virtue of (\ref{h6a}) we obtain:
$$
\p_{\bar t_m}p_i=-m\kappa_m \, \mbox{tr}\, \Bigl (\frac{\p L}{\p x_i}\, L^{-m-1}\Bigr )=
m\kappa_m \, \mbox{tr}\, \Bigl (\frac{\p L^{-1}}{\p x_i}\, (L^{-1})^{m-1}\Bigr )=
\kappa_m \frac{\p }{\p x_i}\, \mbox{tr}\, L^{-m},
$$
which are the Hamiltonian equations
\beq\label{n14}
\p_{\bar t_m}p_i=-\frac{\p \bar H_m}{\p x_i}
\eeq
with $\bar H_m$ given by (\ref{i1a}). In particular, 
\beq\label{n15}
\bar H_1=\frac{\sinh^2(\gamma \eta )}{\gamma^2 \eta^2}\sum_i
e^{-\eta p_i}\prod_{k\neq i}\frac{\sinh (\gamma (x_{ik}-\eta ))}{\sinh (\gamma x_{ik})}.
\eeq

\section{The generating form of equations of motion in higher times}

In the above analysis we parametrized the wave function by residues at its poles.
Another possible parametrization is by zeros and poles. It leads to the so-called
self-dual form of equations of motion. In this section we derive these equations
and show that they provide a generating form of equations of motion 
for the Ruijsenaars-Schneider model in the higher times.

In this section we keep the negative times $\bar {\bf t}$ fixed and consider only the 
dependence on ${\bf t}$. 
In accordance with (\ref{mkp19}) we have $\psi (\mu , x, {\bf t})=\mu^{x/\eta}
e^{\xi ({\bf t}, \mu )}\hat \tau (x, {\bf t})/\tau (x, {\bf t})$ (here
$\hat \tau (x, {\bf t})=\tau (x, {\bf t}-[\mu^{-1}])$), then the auxiliary linear problem
(\ref{mkp13}) acquires the form
\beq\label{g0}
\p_{t_1}\log \frac{\hat \tau (x)}{\tau (x+\eta )}=\mu 
\frac{\hat \tau (x+\eta )\tau (x)}{\tau (x+\eta )\hat \tau (x)}-\mu .
\eeq
For trigonometric solutions $\tau$ is of the form (\ref{ts1}) and for 
$\hat \tau$ we write
\beq\label{g1}
\hat \tau = \prod_i \Bigl (e^{2\gamma x}-e^{2\gamma y_i}\Bigr )=
\prod_i (w-v_i), \quad v_i =e^{2\gamma y_i},
\eeq
parametrizing it by its zeros $y_i$. 
Substituting this into (\ref{g0}), we have:
$$
\sum_i \frac{\dot w_i}{qw-w_i}-\sum_i \frac{\dot v_i}{w-v_i}=
\mu \prod_k \frac{(w-w_k)(qw-v_k)}{(w-v_k)(qw-w_k)}-\mu .
$$
Identifying residues at the simple poles at $w=q^{-1}w_i$ and $w=v_i$, we get the system
of equations
\beq\label{g2}
\left \{ \begin{array}{l}
\displaystyle{2\gamma \dot x_i=\mu (1-q) \prod_{j\neq i}
\frac{w_i-qw_j}{w_i-w_j}\prod_k \frac{w_i-v_k}{w_i-qv_k}}
\\ \\
\displaystyle{2\gamma \dot y_i=\mu (1-q) \prod_{j\neq i}
\frac{qv_i-v_j}{v_i-v_j}\prod_k \frac{v_i-w_k}{qv_i-w_k}}
\end{array}
\right.
\eeq
or
\beq\label{g3}
\left \{ \begin{array}{l}
\displaystyle{\gamma \dot x_i=-\mu \sinh (\gamma \eta )
\prod_{j\neq i}\frac{\sinh (\gamma (x_i-x_j-\eta ))}{\sinh (\gamma (x_i-x_j))}
\prod_{k}\frac{\sinh (\gamma (x_i-y_k))}{\sinh (\gamma (x_i-y_k-\eta ))}}
\\ \\
\displaystyle{\gamma \dot y_i=-\mu \sinh (\gamma \eta )
\prod_{j\neq i}\frac{\sinh (\gamma (y_i-y_j+\eta ))}{\sinh (\gamma (y_i-y_j))}
\prod_{k}\frac{\sinh (\gamma (y_i-x_k))}{\sinh (\gamma (y_i-x_k+\eta ))}}
\end{array}
\right.
\eeq
This is the Ruijsenaars-Schneider analog of the B\"acklund transformation for the 
Calogero-Moser system \cite{W82,ABW09}. These equations appeared in \cite{Suris}
in the context of the integrable time discretization of the Ruijsenaars-Schneider 
model (see also \cite{ZZ18,Z19}). One can show that the equations of motion 
of the Ruijsenaars-Schneider model for 
$x_i$'s follow from (\ref{g3}) and $y_i$'s obey the same equations 
(for the proof see \cite{ZZ18}). 

At the same time these equations contain all the higher equations of motion
in an encoded form.
To see this, we introduce the differential operator
\beq\label{g4}
D(\mu )=\sum_{k\geq 1}\frac{\mu^{-k}}{k}\, \p_{t_k},
\eeq
then $\hat \tau =e^{-D(\mu )}\tau$ and $y_i=e^{-D(\mu )}x_i$. Performing an overall
time shift in the second equation in (\ref{g3}), we can rewrite them in the form
\beq\label{g5}
\left \{ \begin{array}{l}
\displaystyle{\gamma \dot x_i=-\mu \sinh (\gamma \eta )
\prod_{j\neq i}\frac{\sinh (\gamma (x_i-x_j-\eta ))}{\sinh (\gamma (x_i-x_j))}
\prod_{k}\frac{\sinh (\gamma (x_i-e^{-D(\mu )}x_k))}{\sinh 
(\gamma (x_i-e^{-D(\mu )}x_k-\eta ))}}
\\ \\
\displaystyle{\gamma \dot x_i=-\mu \sinh (\gamma \eta )
\prod_{j\neq i}\frac{\sinh (\gamma (x_i-x_j+\eta ))}{\sinh (\gamma (x_i-x_j))}
\prod_{k}\frac{\sinh (\gamma (x_i-e^{D(\mu )}x_k))}{\sinh 
(\gamma (x_i-e^{D(\mu )}x_k+\eta ))}}.
\end{array}
\right.
\eeq
Dividing one equation by the other, we obtain the equations
\beq\label{g6}
\prod_{k=1}^{N}\frac{\sinh (\gamma (x_i -e^{D(\mu )}x_k))}{\sinh 
(\gamma (x_i -e^{D(\mu )}x_k +\eta ))}\,
\frac{\sinh (\gamma (x_i-x_k+\eta ))}{\sinh (\gamma (x_i-x_k-\eta ))}\,
\frac{\sinh (\gamma (x_i -e^{-D(\mu )}x_k -\eta ))}{\sinh 
(\gamma (x_i -e^{-D(\mu )}x_k ))}=-1
\eeq
which are, on one hand, equations of motion for the Ruijsenaars-Schheider system in
discrete time (see \cite{NRK96}) and, on the other, provide the generating form of 
the higher equations of motion in continuous hierarchical times. Indeed, expanding 
(\ref{g6}) in (inverse) powers of $\mu$, one gets the set of the higher equations of motion. 
In particular, equations (\ref{ts16}) are obtained by expansion of (\ref{g6}) up to
$\mu^{-1}$. 

\section{The tau-function for trigonometric solutions}

In this section we prove the determinant formula
for the tau-function of trigonometric solutions
\beq\label{t1}
\tau ' (x, {\bf t}, \bar {\bf t})=\det_{N\times N}\left (wI -\exp \Bigl (\sum_{k\geq 1}
(q^{-k/2}\! -\! q^{k/2}) (t_k L_0^k-\bar t_k L_0^{-k})\Bigr )W_0 \right ), 
\eeq
where $L_0=L(0)$, $W_0 =W(0)$. We recall that the tau-function $\tau'$ is connected with
$\tau$ by formula (\ref{mkp16b}). 

The matrix $\displaystyle{\exp \Bigl (\sum_{k\geq 1}
(q^{-k/2}\! -\! q^{k/2}) (t_k L_0^k-\bar t_k L_0^{-k}) 
\Bigr )W_0}$ can be diagonalized with the help of 
a diagonalizing matrix $V$:
$$
V\exp \Bigl (\sum_{k\geq 1}
(q^{-k/2}\! -\! q^{k/2}) (t_k L_0^k-\bar t_k L_0^{-k})\Bigr )W_0 V^{-1}=W.
$$
There is a freedom in the definition of $V$: it can be multiplied by a diagonal
matrix from the left. We fix this freedom by the condition
\beq\label{t2}
{\bf e}^T W_0^{1/2}={\bf e}^T W^{1/2}V.
\eeq
The matrices $W_0$, $L_0$ satisfy the commutation relation (\ref{ts14}) which we
write here in the form
\beq\label{t3}
q^{-1/2}W_0^{1/2}L_0W_0^{-1/2}-q^{1/2}W_0^{-1/2}L_0 W_0^{1/2}=
W_0^{-1}\dot W_0 {\bf e}\otimes {\bf e}^T.
\eeq
Let us prove, following \cite{Suris}, that the matrices $W$ and $L=VL_0V^{-1}$ satisfy the same commutation 
relation. We have:
$$
q^{-1/2}W^{1/2}LW^{-1/2}-q^{1/2}W^{-1/2}L W^{1/2}=
W^{1/2}(q^{-1/2}L-q^{1/2}W^{-1}LW)W^{-1/2}
$$
$$
=W^{1/2}\Bigl (q^{-1/2}VL_0V^{-1}-q^{1/2}VW_0^{-1}L_0 W_0 V^{-1}\Bigr )W^{-1/2}
$$
$$
=W^{1/2}VW_0^{-1/2}\Bigl (
q^{-1/2}W_0^{1/2}L_0W_0^{-1/2}-q^{1/2}W_0^{-1/2}L_0 W_0^{1/2}\Bigr )
W_0^{1/2}V^{-1}W^{-1/2}
$$
$$
=W^{1/2}VW_0^{-3/2}\dot W_0 \, {\bf e}\otimes {\bf e}^T W_0^{1/2}V^{-1}W^{-1/2}
$$
$$
=W^{1/2}VW_0^{-3/2}\dot W_0 \, {\bf e}\otimes {\bf e}^T.
$$
(The last equality follows from the condition (\ref{t2}).) Denoting
$W^{3/2}VW_0^{-3/2}\dot W_0 \, {\bf e}=\dot W {\bf e}$, we arrive at the 
desired commutation relation.

We are going to prove that the function (\ref{t1}) satisfies the bilinear equations
(\ref{mkp16}), (\ref{mkp16a}) of the 2DTL hierarchy. We begin with equation (\ref{mkp16}):
\beq\label{t4}
\begin{array}{c}
\displaystyle{\mu \frac{\tau (x+\eta , {\bf t}+[\lambda ^{-1}]-[\mu^{-1}])}{\tau (x+\eta ,
{\bf t})}-\lambda \frac{\tau (x , {\bf t}+[\lambda ^{-1}]-[\mu^{-1}])}{\tau (x,
{\bf t})}}
\\ \\
\displaystyle{+(\lambda -\mu )\frac{\tau (x+\eta , {\bf t}+[\lambda ^{-1}])}{\tau (x+\eta ,
{\bf t})}\, \frac{\tau (x , {\bf t}-[\mu^{-1}])}{\tau (x,
{\bf t})}=0}.
\end{array}
\eeq
(Here and below in the proof we put $\bar {\bf t}=0$ and identify $\tau$ and $\tau '$.) 
The similarity transformation with the matrix $V$ under the determinant in (\ref{t1})
allows one to write the following formulas:
\beq\label{t5}
\tau (x, {\bf t})=\det (wI-W),
\eeq
$$
\tau (x, {\bf t}+[\lambda^{-1}])=\det \left (wI-
\frac{\lambda I-q^{1/2}L}{\lambda I-q^{-1/2}L}\, W\right )=
\frac{\det \Bigl (w(\lambda I-q^{-1/2}L)-(\lambda I-q^{1/2}L)W\Bigr )}{\det
\Bigl (\lambda I-q^{-1/2}L \Bigr )}
$$
$$
=\frac{\det \Bigl ((wI-W)(\lambda I -q^{-1/2}L)-\tilde E\Bigr )}{\det
\Bigl (\lambda I-q^{-1/2}L \Bigr )}=
\tau (x, {\bf t})\left (1-\mbox{tr}\, \Bigl (\frac{1}{\lambda I-q^{-1/2}L}\,
\frac{1}{wI -W}\, \tilde E \Bigr )\right ),
$$
where the matrix $\tilde E=W^{-1/2}\dot W E W^{1/2}$ has rank 1 
and we have used the commutation relation
(\ref{ts14}) and the formula $\det (I+A)= 1+\mbox{tr}\, A$ valid for any rank 1 matrix
$A$. Similar calculations yield
$$
\tau (x, {\bf t}-[\mu^{-1}])=
\tau (x, {\bf t})\left (1+\mbox{tr}\, \Bigl (\frac{1}{wI -W}\, 
\frac{1}{\mu I-q^{1/2}L}\,
\tilde E \Bigr )\right ),
$$
$$
\tau (x, {\bf t}+[\lambda^{-1}]-[\mu^{-1}])=\det \left (
wI-\frac{\lambda I-q^{1/2}L}{\lambda I-q^{-1/2}L}\,
\frac{\mu I-q^{-1/2}L}{\mu I-q^{1/2}L}\, W\right )
$$
$$
=\det \left (
wI-\frac{\lambda I-q^{1/2}L}{\lambda I-q^{-1/2}L}\, W
\frac{\mu I-q^{-1/2}L}{\mu I-q^{1/2}L}\right )
$$
$$
=\frac{\det \Bigl (w(\lambda I-q^{-1/2}L)(\mu I - q^{1/2}L)-
(\lambda I-q^{1/2}L)W(\mu I - q^{-1/2}L)\Bigr )}{\det \Bigl (\lambda I-q^{-1/2}L\Bigr )
\det \Bigl (\mu I - q^{1/2}L\Bigr )}
$$
$$
=\frac{\det \Bigl ((\mu I - q^{1/2}L)(wI-W) (\lambda I-q^{-1/2}L)
+(\lambda -\mu )\tilde E\Bigr )}{\det \Bigl (\lambda I-q^{-1/2}L\Bigr )
\det \Bigl (\mu I - q^{1/2}L\Bigr )}
$$
$$
=\tau (x, {\bf t})\left [1+(\lambda -\mu )\, \mbox{tr}\left (
\frac{1}{\lambda I-q^{-1/2}L}\, \frac{1}{wI-W}\,
\frac{1}{\mu I-q^{1/2}L}\, \tilde E\right ) \right ].
$$
Substituting everything into the left hand side of (\ref{t4}), we obtain:
$$
\mbox{LHS of (\ref{t4})} \propto 
q^{-1}\mu \, \mbox{tr} \left [\frac{1}{\lambda I-q^{-1/2}L}\, \frac{1}{wI-q^{-1}W}\,
\frac{1}{\mu I-q^{1/2}L}\, \tilde E\right ]
$$
$$
\phantom{aaaaaaaaaaa}
-\lambda \, \mbox{tr} \left [\frac{1}{\lambda I-q^{-1/2}L}\, \frac{1}{wI-W}\,
\frac{1}{\mu I-q^{1/2}L}\, \tilde E\right ]
$$
$$
+\mbox{tr} \left [ \frac{1}{wI-W}\,
\frac{1}{\mu I-q^{1/2}L}\, \tilde E\right ]-
q^{-1}  \mbox{tr} \left [\frac{1}{\lambda I-q^{-1/2}L}\, \frac{1}{wI-q^{-1}W}\,
\tilde E\right ]
$$
$$
-q^{-1}\mbox{tr} \left [ \frac{1}{wI-W}\,
\frac{1}{\mu I-q^{1/2}L}\, \tilde E\right ]
\mbox{tr} \left [\frac{1}{\lambda I-q^{-1/2}L}\, \frac{1}{wI-q^{-1}W}\,
\tilde E\right ].
$$
This expression is a rational function of $w$ with simple poles at 
$w=w_i$ and $w=q^{-1}w_i$ vanishing at $\infty$. To prove that it actually vanishes
identically it is enough to prove that the residues at the poles are zero. 
The residue at the pole at $w=w_i$ is equal to
$$
-\lambda \sum_{j,k} \Bigl (\frac{1}{\lambda I -q^{-1/2}L}\Bigr )_{ji}
\Bigl (\frac{1}{\mu I -q^{1/2}L}\Bigr )_{ik} w_k^{-1/2}\dot w_k w_j^{1/2}
+\sum_j \Bigl (\frac{1}{\mu I -q^{1/2}L}\Bigr )_{ij}w_j^{-1/2}\dot w_j w_i^{1/2}
$$
$$
-\sum_{j,k,k'}\Bigl (\frac{1}{\mu I -q^{1/2}L}\Bigr )_{ij}
w_j^{-1/2}\dot w_j w_i^{1/2}
\Bigl (\frac{1}{\lambda I -q^{-1/2}L}\Bigr )_{kk'}
\frac{w_{k'}^{-1/2}\dot w_{k'} w_k^{1/2}}{qw_i-w_{k'}}.
$$
Recalling that $\displaystyle{L_{ij}=q^{1/2}\frac{\dot w_i w_i^{-1/2}w_j^{1/2}}{w_i-qw_j}}$,
we can rewrite the last line (the triple sum) in the form
$$
\sum_{j,k}\Bigl (\frac{1}{\mu I -q^{1/2}L}\Bigr )_{ij}
w_j^{-1/2}\dot w_j w_k^{1/2}
\Bigl (\frac{q^{-1/2}L}{\lambda I -q^{-1/2}L}\Bigr )_{ki}
$$
from which it is seen that the residue is zero. The calculation for the residue at
$w=q^{-1}w_i$ is similar. 

The fact that the function (\ref{t1}) is a KP tau-function with respect to the times 
${\bf t}$
follows also from the result of Kasman and Gekhtman \cite{KG01}: for any 
matrices $X$, $Y$, $Z$ such that the matrix $XZ-YX$ has rank $1$ the function
\beq\label{t6}
\tau = \det \left (X\exp \Bigl (\sum_{k\geq 1}t_k Z^k\Bigr )+\exp 
\Bigl (\sum_{k\geq 1}t_k Y^k\Bigr )\right )
\eeq
is a tau-function of the KP hierarchy. In our case $X=-W_0$, $Z=q^{-1/2}L_0$,
$Y=q^{1/2}L_0$ and the condition that
$XZ-YX$ has rank $1$ is equivalent to the 
commutation relation (\ref{ts14}). 

Let us pass to the proof of equation (\ref{mkp16a}) 
which we write here in the equivalent form
\beq\label{t7}
\begin{array}{c}
\displaystyle{\frac{\tau ' (x, {\bf t}+[\lambda ^{-1}], \bar {\bf t}-[\nu ])}{\tau ' (x-\eta ,
{\bf t}, \bar {\bf t})}-
\frac{\nu}{\lambda}\,
\frac{\tau ' (x+\eta , {\bf t}+[\lambda ^{-1}], \bar {\bf t}-[\nu ])}{\tau ' (x ,
{\bf t}, \bar {\bf t})}}
\\ \\
\displaystyle{-\Bigl (1-\frac{\nu}{\lambda}\Bigr )
\frac{\tau ' (x, {\bf t}, \bar {\bf t}-[\nu ])}{\tau ' (x-\eta ,
{\bf t}, \bar {\bf t})}\,
\frac{\tau ' (x , {\bf t}+[\lambda ^{-1}])}{\tau ' (x ,
{\bf t}, \bar {\bf t})}=0.}
\end{array}
\eeq
The calculations similar to the ones done above yield:
$$
\tau '(x, {\bf t}+[\lambda^{-1}], \bar {\bf t})=
\tau '(x, {\bf t}, \bar {\bf t})
\left (1-\mbox{tr}\, \Bigl (\frac{1}{\lambda I-q^{-1/2}L}\,
\frac{1}{wI -W}\, \tilde E \Bigr )\right ),
$$
$$
\tau '(x, {\bf t}, \bar {\bf t}-[\nu ])=q^N \tau '(x-\eta , {\bf t}, \bar {\bf t})
\left (1+\mbox{tr}\, \Bigl (\frac{q}{wI -qW}\,\frac{1}{\nu I-q^{1/2}L}\,
\tilde E \Bigr )\right ),
$$
$$
\tau '(x, {\bf t}+[\lambda ^{-1}], 
\bar {\bf t}-[\nu ])=q^N \tau '(x-\eta , {\bf t}, \bar {\bf t})\phantom{aaaaaaaaaaaaaa}
$$
$$
\phantom{aaaaaaaaaaaaaaaaaa}
\times \left (1+(\lambda -\nu )\, \mbox{tr}\left (\frac{1}{\lambda I-q^{-1/2}L}\,
\frac{q}{wI -qW}\,\frac{1}{\nu I-q^{1/2}L}\,
\tilde E \right )\right ).
$$
Using the results of the above calculation, it
is not difficult to see that the substitution into (\ref{t7}) gives the identity,
so the equation (\ref{mkp16a}) is proved.

\section{Conclusion}

The main result of this paper is establishing the precise correspondence between
trigonometric solutions of the 2D Toda lattice hierarchy and the hierarchy of the Hamiltonian
equations for the integrable Ruijsenaars-Schneider model with higher Hamiltonians. 
The zeros of the tau-function move as particles of the Ruijsenaars-Schneider model.
We have shown that the $m$th time flow $t_m$ of the 2DTL hierarchy gives rise to the 
flow with the Hamiltonian $H_m$ of the Ruijsenars-Schneider model 
proportional to $\mbox{tr}\, L^m$, where $L$ is the Lax matrix, while the time flow
$\bar t_m$ gives rise to the Hamiltonian flow with the Hamiltonian $\bar H_m$ 
proportional to $\mbox{tr}\, L^{-m}$. In some sense this correspondence is 
simpler and more natural then a similar correspondence between the KP hierarchy and 
trigonometric Calogero-Moser hierarchy \cite{Z19a}, which in principle can be obtained
from our results in the limit $\eta \to 0$.

\section*{Acknowledgments}

The work of V.P. was supported in part by RFBR grant 18-01-00273.
The research of A.Z. was carried out within the HSE University Basic 
Research Program and funded (jointly) by the Russian Academic Excellence Project '5-100'.  
The work of A.Z. was also supported in part by RFBR grant
18-01-00461.

\end{document}